\documentclass[twocolumn,showpacs,preprintnumbers,amsmath,amssymb]{revtex4}

\usepackage{epsfig}
\usepackage{graphicx}
\usepackage{dcolumn}
\usepackage{natbib}
\usepackage{soul}
\usepackage[dvips]{color}

\usepackage{amsmath,amssymb}

\newcommand{\br}{\boldsymbol{r}}
\newcommand{\bu}{\boldsymbol{u}}

\newcommand{\bxi}{\boldsymbol{\xi}}
\newcommand{\Kn}{\mbox{Kn}}

\begin{document}

\title{A finite-difference lattice Boltzmann approach for gas microflows}

\author{G. P. Ghiroldi}
\author{L. Gibelli}
\email[Corresponding author: ]{livio.gibelli@polimi.it}
\affiliation{Dipartimento di Scienze e Tecnologie Aerospaziali,
             Politecnico di Milano, 
             20156 Milano, Italy}

\date{\today}

\begin{abstract}
Finite-difference Lattice Boltzmann (LB) models are proposed for simulating gas flows in devices with microscale geometries.
The models employ the roots of half-range Gauss-Hermite polynomials as discrete velocities.
Unlike the standard LB velocity-space discretizations based on the roots of full-range Hermite polynomials, 
using the nodes of a quadrature defined in the half-space permits a consistent treatment of kinetic boundary conditions.
The possibilities of the proposed LB models are illustrated by studying the one-dimensional Couette flow and the two-dimensional driven cavity flow. 
Numerical and analytical results show an improved accuracy in finite Knudsen flows as compared with standard LB models.
\end{abstract}

\pacs{47.11.-j, 47.15.-x, 47.61.-k}

\maketitle

  
The outgrowing development of {micro\-electro\-me\-cha\-ni\-cal systems} (MEMS)
has spurred interest in studying gas flows in devices with microscale 
geometries.
These flows, which are referred to as gas microflows, are usually distinguished by relatively large Knudsen numbers 
and small Mach numbers. 
Because of the large Knudsen numbers, the conventional hydrodynamic approach breaks down and 
a description based on the Boltzmann equation is required. The low Mach 
numbers permit to replace the collision integral of the Boltzmann equation with a simpler kinetic model, such as the one 
proposed by Bathnagar, Gross and Krook and, independently, by Welander (BGKW), as well as to linearize the resulting 
kinetic equation around the equilibrium state~\cite{c88}.
Numerous studies on gas microflows based on the numerical solution of the linearized BGKW kinetic model equation~\cite{lgffc07,nv07,aczgkp11} and/or
the linearized Boltzmann equation~\cite{cffggl08} have been reported over the years.
Recently, the Lattice Boltzmann (LB) method has attracted considerable 
interest as an alternative tool for studying gas flows in microfluidic devices~\cite{zgbe06,iri11,kpb08,mz11a,sbys11,mzs11,mz11b}.
Although it was evolved from lattice-gas cellular automaton models for 
mimicking the Navier-Stokes hydrodynamics, the LB method can
potentially describe gas microflows since it can be viewed
as the discrete ordinate method to solve the linearized 
BGKW kinetic model equation~\cite{hl87,sh98}.
By using the same notation as in Ref.~\cite{mz11b},
the non dimensional form of LB models in the absence of external force fields reads
\begin{equation}
\label{eq:BGKW_lattice}
\frac{\partial f_\alpha}{\partial t} + {\bf \bxi_\alpha} \cdot \nabla{f_\alpha} = -\frac{1}{\Kn} \left( f_\alpha -f_\alpha^{eq} \right), 
\end{equation}
where the Knudsen number is defined as $\Kn=\mu_0 \sqrt{RT_0}/(p_0 l)$ with $\mu_0, T_0, p_0$ the reference gas viscosity, temperature and 
pressure, respectively, $l$ the characteristic length and 
\begin{eqnarray}
\label{eq:f_alpha}
 f_\alpha (\br,t)= w_\alpha f(\br,\bxi_\alpha,t)/\omega(\bxi_\alpha), \\
\label{eq:feq_alpha}
 f^{eq}_\alpha (\br,t)= w_\alpha f^{eq}(\br,\bxi_\alpha,t)/\omega(\bxi_\alpha),
\end{eqnarray}
being
\begin{equation}
 \omega(\bxi) = \frac{1}{\left(2\pi\right)^{d/2}} \exp{\left(-\frac{\xi^2}{2}\right)}.
\end{equation}
In Eqs.~\eqref{eq:f_alpha} and~\eqref{eq:feq_alpha}, $f$ is the distribution function, $f^{eq}$ is the $d$-dimensional equilibrium Maxwellian
and $w_\alpha, \bxi_\alpha$ are the $n$ weights and nodes determined from a quadrature formula, respectively~\cite{syc06}.
Since we are considering isothermal flows, the terms related to the temperature in the second-order approximation of the equilibrium Maxwellian can be 
disregarded and $f^{eq}$ can thus be written as
\begin{equation}
\label{eq:Maxwellian}
 f^{eq} (\br,t)=\omega(\bxi) \rho(\br,t) \left[1+\bxi\cdot \bu(\br,t) \right],
\end{equation}
where $\rho$ and $\bu$ are the density and macroscopic velocity, respectively, which can be computed by sums over the discrete velocity set

\begin{eqnarray}
 \label{eq:density}
 \rho (\br,t) & = & \sum_{\alpha=1}^{n} f_\alpha (\br,t), \\
 \label{eq:velocity}
 \rho (\br,t) \bu (\br,t) & = & \sum_{\alpha=1}^{n} \bxi_\alpha f_\alpha (\br,t).
\end{eqnarray}
Likewise, boundary conditions can be derived by a direct discretization of Maxwell's diffuse-specular scattering kernel~\cite{ak02}.
For the sake of simplicity, let us consider a plane wall at $y=0$ with reference temperature and constant velocity $u_w$ in the positive $x$-direction.
Let us further suppose that the gas fills the half space $y>0$ and molecules which strike the wall are re-emitted according to Maxwell's
scattering kernel with complete accommodation. The discrete form of the kinetic boundary condition at a point $\br_w$ of the solid surfaces reads 

\begin{equation}
\label{eq:BC}
f_{\alpha}(\br_w,t) = \omega(\bxi_\alpha) \rho_{w} (\br_w,t) \left( 1+\xi_{x,\alpha} u_w \right), 
\mbox{\hspace{0.2cm}} \xi_{y,\alpha}>0.
\end{equation}
In Eq. (\ref{eq:BC}), $\rho_{w}$ can be obtained through the {im\-per\-mea\-bi\-li\-ty} condition, which states that the normal
component of the gas velocity on the wall vanishes

\begin{equation}
\label{eq:rho}
\rho_{w} (\br_w,t) = -(2\pi)^{1/2} \sum_{\xi_{y,\alpha}<0} \xi_{y,\alpha} f_\alpha (\br_w,t).
\end{equation}
For two-dimensional flows and very low Knudsen numbers, the $D_{2}Q_{9}$ model provides accurate results.
We here use the standard terminology and denote by $D_mQ_n$ the $m$ dimensional LB models with $n$ discrete velocities.  
As $\Kn$ increases, high-order LB models are needed 
to correctly reproduce non equilibrium effects, such as the velocity slip at
the solid walls and the nonlinear stress-strain relationship within the 
Knudsen layer~\cite{kpb08}. 
In the framework of single relaxation time modeling, several high-order LB models have been developed.
Some have been derived by using a local mean free path in order to account for the presence of solid surfaces~\cite{zgbe06}. 
Although these models have been shown to be effective in many applications,
they are phenomenological in nature and, as such, not perfectly general.
Composite models have also been developed which result from the superposition of discrete velocities determined from odd and even quadrature 
formula~\cite{iri11}. 
However, the most common strategy is using a greater number of discrete velocities determined from a full-range 
Gauss-Hermite quadrature~\cite{kpb08,mz11a}, possibly adopting a multiscale approach to contain the increase in the computational cost~\cite{mzs11}.
However, more 
discrete velocities do not guarantee an improved accuracy~\cite{kpb08,mz11a,mz11b}.
It has been demonstrated that this is due to the quadrature effect in dealing with the boundary conditions~\cite{sbys11}.
As a matter of fact, abscissae of the full-range Gauss-Hermite quadrature schemes  
are derived to obtain accurate evaluation of the moments of the distribution function
defined over the entire velocity space. In contrast, they provide 
only an approximately estimate of the half-range integrals that enter in the formulation of
kinetic boundary conditions~\cite{ak02}.

\begin{table}[t!]
\caption{\label{tab:nodes_weights}
         Nodes, $\xi_\alpha$, and weights, $w_\alpha$, of the one-dimensional half-range Gauss-Hermite quadrature.}
\begin{ruledtabular}
\begin{tabular}{ccc} 
Quadrature & $\xi_\alpha$ & $w_\alpha$ \\  \hline
$D_{1}Q^{h}_{4}$ & $\begin{array}[t]{c}  
               \mp 0.4245383286 \\
               \mp 1.77119083
              \end{array}$
            & $\begin{array}[t]{c} 
              0.3613798911 \\ 
              0.1386201089                      
              \end{array}$ \\
$D_{1}Q^{h}_{6}$ & $\begin{array}[t]{c}  
              \mp 0.2694842630 \\
              \mp 1.199609295 \\
              \mp 2.545268446
              \end{array}$
            & $\begin{array}[t]{c} 
              0.2516453504 \\
              0.2236832664 \\
              0.0246713831 \\                     
              \end{array}$ \\ 
$D_{1}Q^{h}_{8}$ & $\begin{array}[t]{c}  
              \mp 0.1891884657 \\
              \mp 0.8829284442 \\
              \mp 1.898635201 \\
              \mp 3.199890790 \\
              \end{array}$
            & $\begin{array}[t]{c} 
              0.1835325640 \\
              0.2375842404 \\
              0.07528686870 \\
              0.003596326913 \\                     
              \end{array}$    
\end{tabular}
\end{ruledtabular}
\end{table}

\noindent
In this Rapid Communication, we want to show that non-equilibrium gas flows can be more
accurately described by using a discrete velocity set different from the one employed 
by standard finite-difference LB models. 
More specifically, we propose finite-difference LB models which use the roots of half-range Hermite polynomials as quadrature nodes.
Table~\ref{tab:nodes_weights} gives nodes and weights of the one-dimensional half-range Gauss-Hermite quadrature. The two-dimensional
quadrature can be obtained from the tensor product of the corresponding one-dimensional quadrature.
Notice should be made that the roots of half-range Hermite polynomials are irrational and therefore the resulting fully-discrete numerical
schemes are computationally more demanding than the simpler ``stream-and-collide'' algorithm. 
The exact space discretization of the advection step of on-lattice LB models 
is therefore no longer possible and the potential high efficiency of their parallel implementations is partially lost.
However, off-lattice schemes may offer some advantages such as enhanced geometrical flexibility and, as shown in 
the present work, the capability of describing rarefaction effects. The use of high performance computing is still feasible but
the porting of algorithms involving finite-difference approximations certainly requires some additional effort~\cite{fgg11}.
In comparison with alternative finite-difference LB models, the proposed velocity space discretization permits to explicitly account for the discontinuity of 
the distribution function and therefore leads to a faster convergence of the solution close to 
solid surfaces. In kinetic theory applications, the importance of a consistent treatment of boundary conditions has been recognized as early as 
the sixties of the past century~\cite{gjz57} and the half-range discrete ordinate method has been widely used since then~\cite{hg67,bcrs01,fgf09,g12,gg13}. 
By contrast, in LB simulations of gas flows in microchannels, most of high-order LB models use the roots of full-range Hermite polynomials as 
discrete velocities~\cite{zgbe06,iri11,mzs11}.
Although it is well known that finite-difference LB models can be developed from different quadrature formula~\cite{syc06}, 
to the authors' knowledge, no previous works have pointed out that the quadrature based on half-range Hermite polynomials can easily address
the issue of boundary conditions for the LB simulations of gas microflows.  
An important correspondence can be identified between the approach developed in the present work 
and the moment method presented in Ref.~\cite{fgf09} for studying gas microflows. There, the 
isothermal linearized BGKW kinetic model equation has been solved by expanding the distribution function as a series of
half-range Hermite polynomials. 
Expansion coefficients are the moments of the distribution function which, in turn, are strictly related to the macroscopic quantities.
By using the half-range Gauss-Hermite quadrature formula for evaluating these integrals, a one-to-one correspondence can be identified 
between the expansion coefficients and the values of the distribution function at the roots of half-range Hermite polynomials.
The two approaches are therefore equivalent even though, from the computational standpoint, the formulation which is here proposed is more efficient. 
It is worth noticing that already in Ref.~\cite{ak02} it has been pointed out that 
the accuracy in dealing with boundary conditions can be improved by evaluating integrals which enter in their
definition by means of a quadrature formula defined in the half-space.
However, this possibility has not been further developed 
because of the mismatch between the nodes of the quadrature used at the boundary and those in the bulk.
In the present work, however, the nodes of half-range Gauss-Hermite quadrature formula are used in the whole domain.
\begin{figure}[t!]
\begin{center}
\includegraphics[scale=0.3]{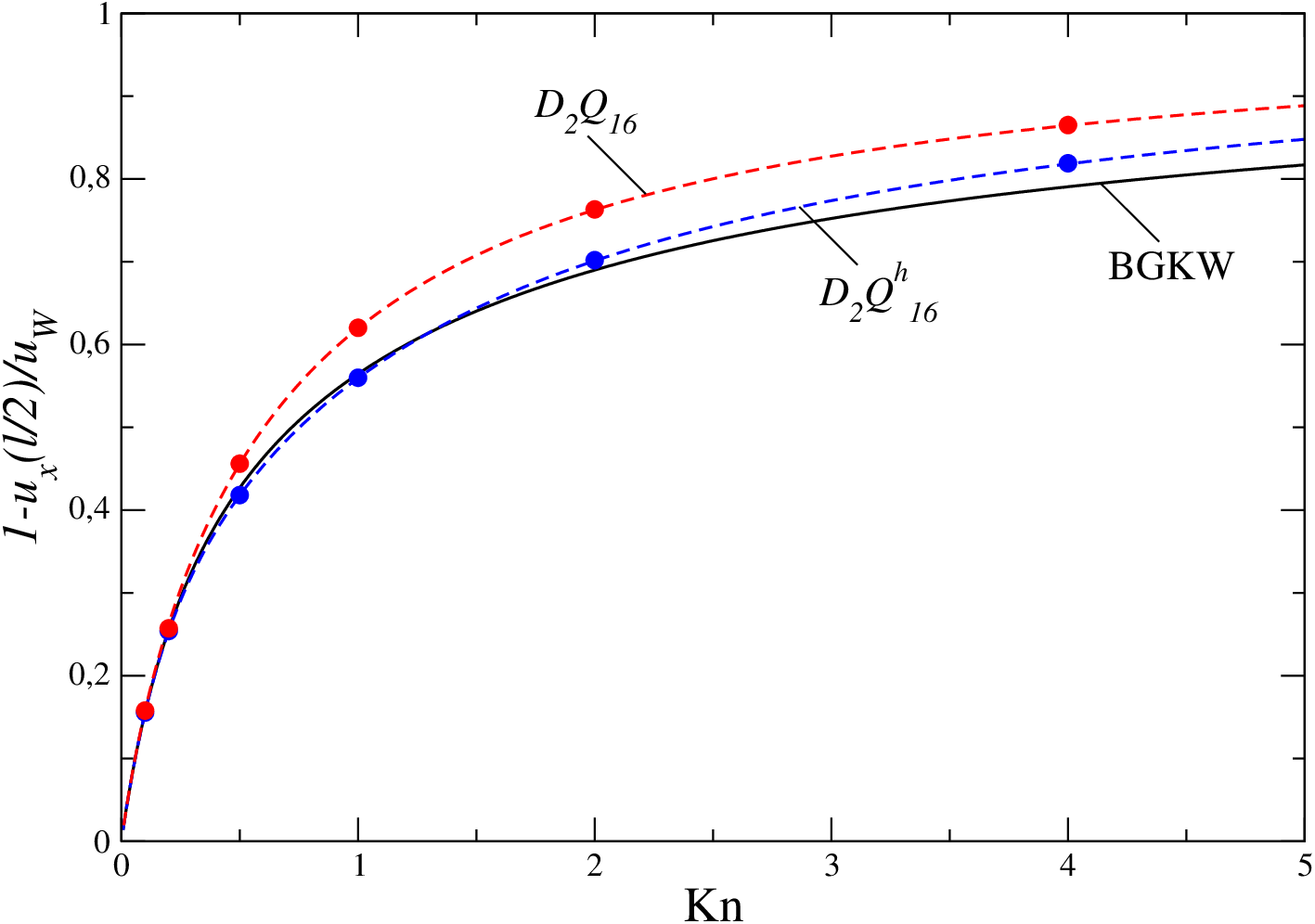}
\end{center}
\caption{Velocity slip at the upper plate versus the Knudsen number.}
\label{fig:slip}
\end{figure}
\begin{figure}[t!]
\vspace{0.5cm}
\begin{center}
\includegraphics[scale=0.3]{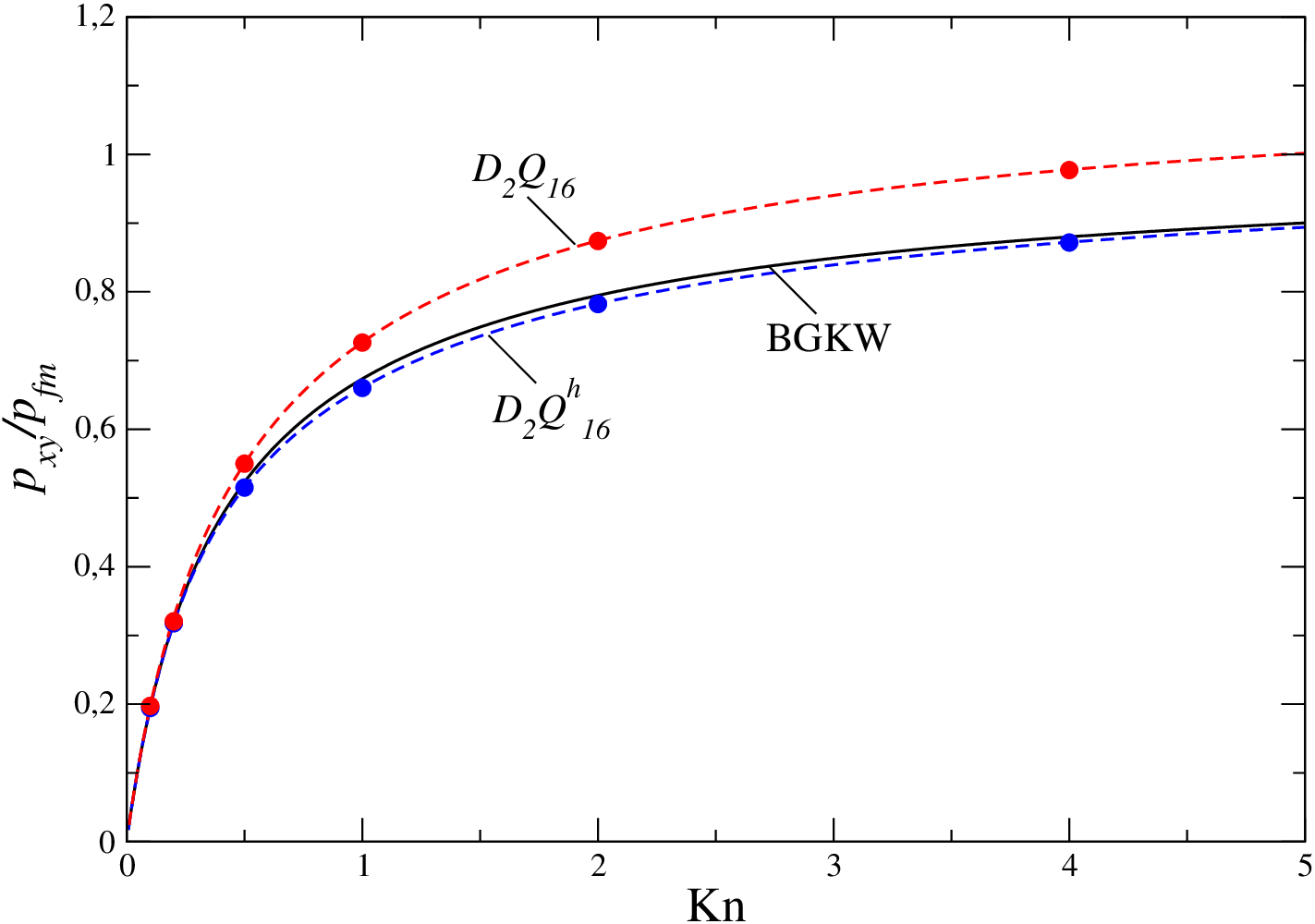}
\end{center}
\caption{Reduced $xy$-component of the stress tensor versus the Knudsen number.}
\label{fig:pxy}
\end{figure}
\noindent
We demonstrate, both analytically and numerically, that the proposed approach can be used to simulate gas microflows by
studying two classical driven boundary value problems, i.e., the Couette and the cavity flows.
Both problems are regarded as two-dimensional and the driven velocity is assumed sufficiently low so 
that the gas flow can be considered in the linearized regime~\cite{ggdi12}.
A fully discrete numerical scheme is derived by a first-order time splitting of the evolution operator and couples a first-order upwind scheme for the transport 
step with a first-order explicit Euler scheme for the relaxation step. 
More sophisticated methods should be used for both coupling and solving transport and relaxation steps if a more accurate and efficient numerical scheme is needed.
However, the main aim of the work is to show that an improved accuracy in describing finite Knudsen flows can be 
achieved by using the roots of half-range Hermite polynomials as discrete velocities. 
This has been assessed by running the same numerical code twice, once with nodes and weights as given by full-range Gauss-Hermite quadrature and 
once with nodes and weights as given by half-range Gauss-Hermite quadrature.
The validity of this comparison is therefore independent on the numerical scheme that has been used to solve the discrete kinetic equation. \\
In the Couette flow problem, the upper plate moves in the $x$ direction with a velocity $u_w$ and the lower plate moves with a velocity $-u_w$.
Diffuse boundary conditions have been implemented at the plates and periodic boundary conditions have been utilized at opposite ends of 
the channels. The two plates are separated by the distance $l$.
It is worth noticing that the proposed models employ the quadrature nodes of the positive and negative half-range Gauss-Hermite quadrature formula as discrete
velocities both in the $x$- and $y$-directions. However numerical computations showed that the same results are obtained if 
the $x$-direction of the velocity space is discretized by using the roots of full-range Hermite polynomials.
By proceeding as in Ref.~\cite{fgf09}, it is not difficult to obtain a closed-form solution of the $D_{2}Q^{h}_{16}$ LB model for the stationary planar Couette 
flow

\begin{widetext}
\begin{eqnarray}
\label{eq:velocity_exact}
u_x(y/l) & = & u_{w}
         \frac{\left[\sinh{\left(\frac{0.9494}{\Kn}\right)}+0.7978 \cosh{\left(\frac{0.9404}{\Kn}\right)}\right]\frac{y}{l}+
         0.5642 \Kn \sinh{\left(\frac{1.881}{\Kn}\frac{y}{l}\right)}}
         {(0.7071 \Kn +0.3989)\cosh{\left(\frac{0.9404}{\Kn}\right)}+(1.128 \Kn +0.5)\sinh{\left(\frac{0.9404}{\Kn}\right)}}, \\
\label{eq:pressure_exact}
\frac{\sqrt{RT_0}p_{xy}}{u_w p_0} & = & -\Kn \frac{\sinh{\left(\frac{0.9494}{\Kn}\right)}+0.7979\cosh{\left(\frac{0.9404}{\Kn}\right)}} 
                 {(0.7071 \Kn +0.3989)\cosh{\left(\frac{0.9404}{\Kn}\right)}+(1.128 \Kn +0.5)\sinh{\left(\frac{0.9404}{\Kn}\right)}}.
\end{eqnarray}
\end{widetext}
Beside their intrinsic interest, Equations~\eqref{eq:velocity_exact} and~\eqref{eq:pressure_exact}, as well as a similar closed-form solution
obtained in Ref.~\cite{akaap07} for the $D_{2}Q_{16}$ model, are of practical importance in that they permit to validate the numerical code. 
Figures~\ref{fig:slip} and~\ref{fig:pxy} show the velocity slip, $1-u_{x}/u_{w}$, at the upper plate
and the ratio between the $xy$-component of the stress tensor, $p_{xy}$, and the value of the pressure in the free molecular regime, 
$p_{fm}=-2p_0 u_{w}/\sqrt{\pi R T_0}$, versus the Knudsen number, $\Kn$, respectively.
Both Figures use the same conventions.
Solid line are the solutions of the linearized BGKW kinetic model equation obtained by means of the moment method described in Ref.~\cite{fgf09}.
Dashed lines are the closed-form solutions of the $D_{2}Q^{h}_{16}$ model, as given by Eqs.~\eqref{eq:velocity_exact} and~\eqref{eq:pressure_exact} 
evaluated at $y/l=1/2$, and of
the $D_{2}Q_{16}$ model, which is reported in Ref.~\cite{akaap07}. Solid circles are the numerical solutions of the $D_{2}Q^{h}_{16}$ and $D_{2}Q_{16}$ models. 
Only even-order LB models have been considered since they perform significantly better than those with an odd-order
quadrature no matter how the order is~\cite{kpb08,mz11a}.
As it is clearly shown, although the same number of discrete velocities is employed, the $D_{2}Q^{h}_{16}$ model gives much improved results as compared with standard 
$D_{2}Q_{16}$ model. The prediction of the velocity slip shows a very good match with the results obtained by solving the linearized BGKW kinetic model equation not only 
in the continuum and slip flow regimes but also in the early transition regime.
The $xy$-component of the stress tensor shows a quite good agreement even in a wider range of Knudsen numbers. 
\begin{figure}[t!]
\begin{center}
\includegraphics[scale=0.3]{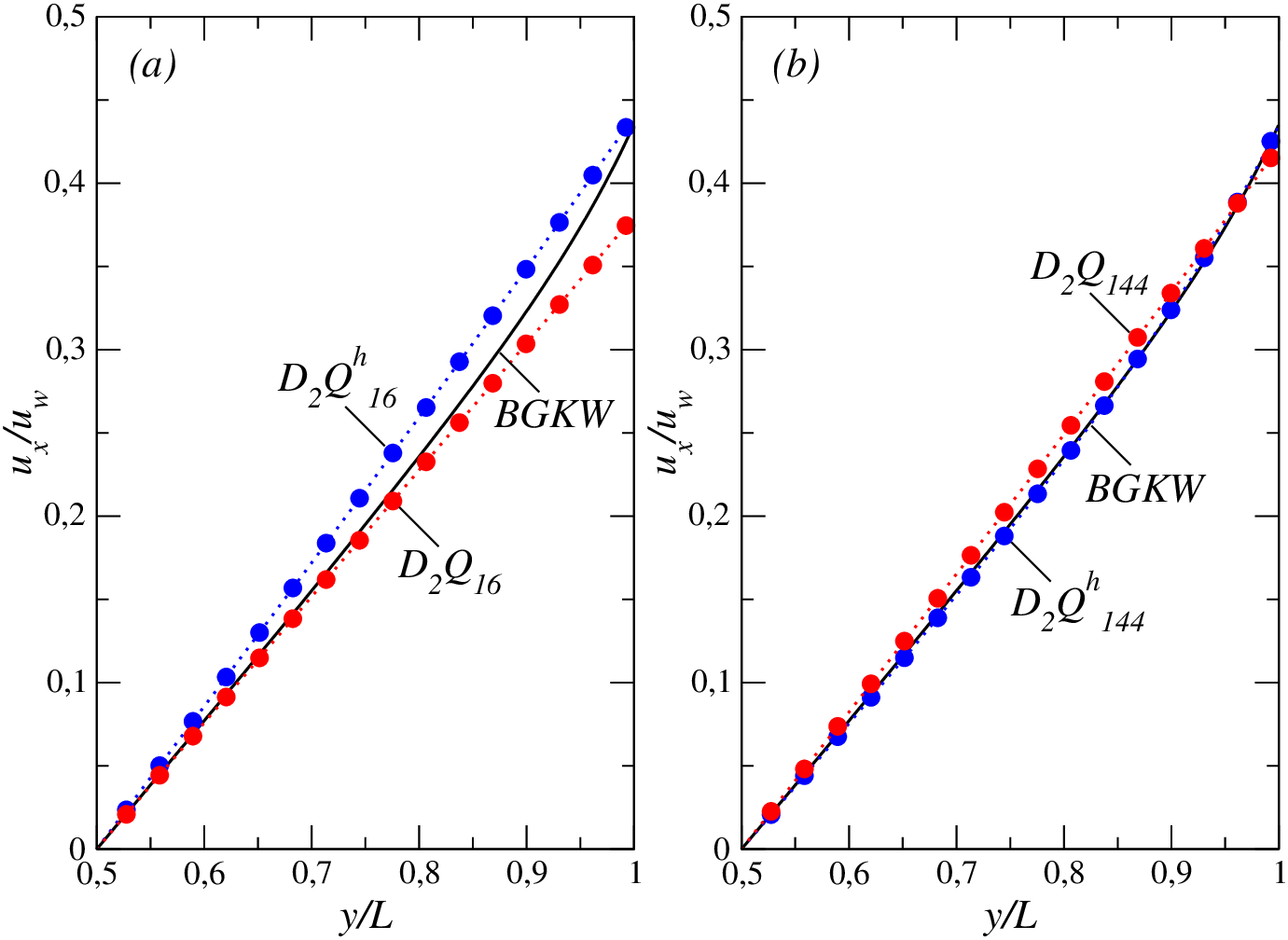}
\end{center}
\caption{Nondimensional velocity profile through the half-channel. $\Kn=1$. (a) $n=16$, (b) $n=144$.}
\label{fig:velocity}
\end{figure}
Convergence of the results at $\Kn=1$ with increasing number of discrete velocities is reported in Fig.~\ref{fig:velocity} for the velocity profile.
Solid line is the solution of the linearized BGKW kinetic model equation obtained by means of the moment method described in Ref.~\cite{fgf09}. 
Dotted lines with solid circles are the numerical solutions obtained by using $n=16$ (left panel) and $n=144$ (right panel) discrete velocities, respectively.
The $D_2Q_{16}$ model provides a slightly better description of the gas behavior in the bulk of the flow.
A possible explanation is that the full-range Gauss-Hermite quadrature with $n=16$ discrete velocities integrates exactly continuous polynomials of order five 
in each velocity component whereas the half-range Gauss-Hermite quadrature with the same
number of nodes integrates exactly linear polynomials with a possible discontinuity at the origin. It is thus reasonable that the former can be more accurate
in the bulk of the flow where the distribution function is continuous whereas the latter can perform better close to the solid surface where the
distribution function is expected to be discontinuous.
Nevertheless, as results with $n=144$ clearly show, using the roots of half-range Hermite polynomials as discrete velocities greatly speed-up 
the convergence to the kinetic theory solution. A similar analysis for the $xy$-component of the stress tensor shows that the numerical prediction of the 
$D_{2}Q^{h}_{16}$ model at $\Kn=1$ is affected by a relative error which is lower than $5\%$ and at least $12$ nodes per each velocity component should be used for 
the standard high-order LB models to achieve the same level of accuracy.
\begin{figure}[t!]
\vspace{0.5cm}
\begin{center}
\includegraphics[scale=0.3]{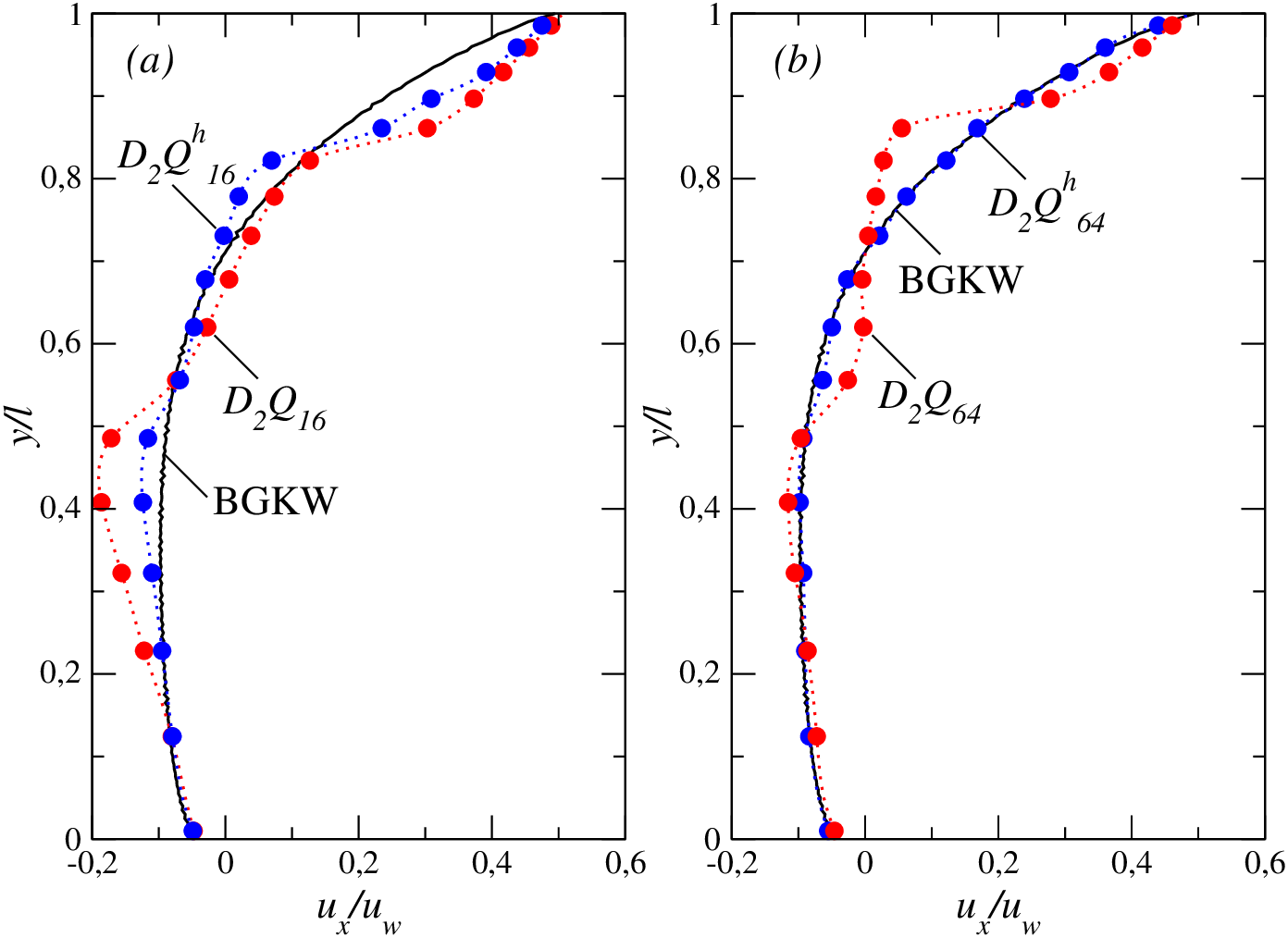}
\end{center}
\caption{Profile of the nondimensional horizontal component of the mean velocity crossing the center of the cavity. $\Kn=0.05$. (a) $n=16$, (b) $n=64$.} 
\label{fig:comparison_cavity}
\end{figure}
The possibilities of the proposed models are further illustrated by solving the square driven cavity flow problem~\cite{nv05}.
All the walls have a length $l$ and are fixed and isothermal.
The flow is driven by the uniform translation of the top. 
Numerical simulations have been carried out for Knudsen numbers in the range $[0.05,0.4]$.
This problem constitutes a severe test to assess the capability of the proposed velocity space discretization to describe the gas behavior in complex geometries.
Indeed, the discontinuities which exist at the four corners, particularly the two at the top, propagate inside the computational domain and might cause 
the quadrature to fail. Since the discontinuities decay with distance owing to molecular collisions, it is expected that these numerical problems become more severe 
as the Knudsen number is increased. 
In Fig.~\ref{fig:comparison_cavity}, the profile of the nondimensional horizontal component of 
the macroscopic velocity, $u_{x}/u_w$, crossing the center of the cavity are thus show for the higher Knudsen number we considered, $\Kn=0.4$. 
Solid line is the solution of the linearized BGKW equation obtained with the numerical method described in Ref.~\cite{nv05}.
Dotted lines with solid circles are the numerical solutions provided by full- and half-range LB models with $n=16$ (left panel) and $n=64$ (right panel)
discrete velocities. As it was clearly shown, both methods suffer from an unphysical oscillatory behavior due to the difficulty of the quadratures to
evaluate a region where discontinuities are present. Although more sophisticated methods can be adopted~\cite{nv05}, 
we here simply notice that the problem can be overcome by a reasonable increase in the number of discrete velocities.
As for the case of the Couette flow, the proposed approach shows better convergence properties. For instance, the errors in the ${\cal L}^1$-norm
of full- and half-range LB models using $n=16$ discrete velocities are $0.0445$ and $0.0280$, respectively,
and reduce to $0.0218$ and $0.00354$ for $n=64$. 
Table~\ref{tab:DeG} reports a comparison of the values of the mean dimensionless shear stress along the moving plate, $D$, and 
the dimensionless flow rate of the main vortex, $G$, obtained by the different models~\cite{nv05}. The predictions of the $D_{2}Q^{h}_{16}$ model
show a better agreement with the kinetic theory results than those of the $D_{2}Q_{16}$ model. 
In spite of the unphysical oscillations which affect the macroscopic velocity,
the drag coefficient converges up to two significant figures.
Instead the error in the reduced flow rate is less than $1\%$ for $\Kn=0.05$ but rapidly increases for greater Knuden numbers. 
However, as shown in Ref.~\cite{fgf09}, 
a quite good agreement can be found up to $\Kn=10$ if the $D_{2}Q^{h}_{36}$ is used.   

\begin{table}[h!]
\caption{\label{tab:DeG}
         Drag coefficient, $D$, and reduced flow rate, $G$, versus the Knudsen number, $\Kn$.}
\begin{ruledtabular}
\begin{tabular}{ccccccc} 
      & \multicolumn{3}{c}{$D$}                   & \multicolumn{3}{c}{$G$}                   \\  \cline{2-4} \cline{5-7}
$\Kn$ & BGKW  & $D_{2}Q^{h}_{16}$ & $D_{2}Q_{16}$ & BGKW  & $D_{2}Q^{h}_{16}$ & $D_{2}Q_{16}$ \\ \hline 
0.05  & 0.258 & 0.256             & 0.270         & 0.154 & 0.153             & 0.162 \\
0.1   & 0.328 & 0.327	          & 0.350	  & 0.136 & 0.128	      & 0.153 \\
0.2   & 0.425 & 0.417	          & 0.467         & 0.121 & 0.109 	      & 0.148 \\
0.4   & 0.385 & 0.384	          & 0.421	  & 0.110 & 0.0961	      & 0.145 \\ 
\end{tabular}
\end{ruledtabular}
\end{table}

\noindent
To summarize, we have shown that, in comparison with full-range finite-difference high-order LB models,
using the nodes of half-range Gauss-Hermite quadrature as discrete velocities 
permits to consistently deal with kinetic boundary conditions and thus to achieve a more
accurate description of the gas behavior close to solid surfaces.
Applications to one- and two-dimensional driven boundary value problems show that, even using a small number of discrete velocities,
accurate results can be obtained in a wide range of Knudsen numbers which extends up to the early transition regime.
The proposed approach is also of interest in that, as it can be deduced from Ref.~\cite{fgf09}, half-range LB models with $N$ discrete velocities 
simplify 
to full-range LB models with $N/2^d$ velocities far away from solid surfaces, being $d$ the dimension of the physical space. 
This suggest the development of a hybrid 
approach based on their coupling which combines the capability of half-range LB models to accurately describe the gas behavior in the Knudsen layers with
the higher computational efficiency of full-range LB models when applied to the bulk of the gas.

\begin{acknowledgments}
The authors would like to thank Prof. A. {Frezzot\-ti} for critically reading the paper and Prof. D. {Va\-lou\-ge\-or\-gis} for providing his numerical code.
This work has been partly supported by ``Progetto Giovani Ricercatori GNFM 2013'', 
Regione Lombardia and CILEA Consortium through a LISA Initiative (Laboratory for Interdisciplinary Advanced
Simulation) 2011 grant [link:http://lisa.cilea.it].
\end{acknowledgments}

\end{document}